\documentclass{cpcauth}

\usepackage{amssymb}

\begin{document}

\begin{frontmatter}

% Title, authors and addresses

% use the thanksref command within \title, \author or \address for footnotes;
% use the corauthref command within \author for corresponding author footnotes;
% use the ead command for the email address,
% and the form \ead[url] for the home page:
% \title{Title\thanksref{label1}}
% \thanks[label1]{} % \author{Name\corauthref{cor1}\thanksref{label2}}
% \ead{email address} % \ead[url]{home page} % \thanks[label2]{}
% \corauth[cor1]{} % \address{Address\thanksref{label3}} 
% \thanks[label3]{}

\title{Fortran MPI Checkerboard Code for SU(3) Lattice Gauge Theory II} 

% use optional labels to link authors explicitly to addresses:
% \author[label1,label2]{} % \address[label1]{} % \address[label2]{}

\author{Bernd A. Berg$^{\,a}$ and Hao Wu$^{\,a}$ }

\address{~~\\
$^{a)}$ Department of Physics, Florida State University, 
Tallahassee, FL 32306-4350, USA}

\date{\today} % \date{December xxx 2008}

\begin{abstract}
% \narrowtext
We study the performance of MPI checkerboard code for SU(3) lattice 
gauge theory as function of the number of MPI processes, which run 
in parallel on an identical number of CPU cores. Computing platforms 
explored are a small PC cluster at FSU and the Cray at NERSC.
\end{abstract}

\end{frontmatter}

\section{Introduction} \label{sec_intro}

In a previous paper \cite{Be09} Fortran MPI checkerboard code for 
Markov Chain Monte Carlo (MCMC) simulations of pure SU(3) Lattice 
Gauge Theory (LGT) with the Wilson action is introduced and a number 
of tests and verifications are provided. These programs allow for 
simulations with periodic boundary conditions (PBC) as well as for
the geometry of a double-layered torus (DLT), which remains to be 
explored in more details. 

Here we extend this work and investigate the performance as function 
of the number of CPU cores used by an equal number of MPI processes. 
Tests were carried out on a cluster of 2 high end PCs with together 
16 cores at the High Energy Physics (HEP) group of the Florida State 
University (FSU) and on a Cray XT4 (named Franklin) with 38$\,$640 
cores at the National Energy Research Scientific Computing Center 
(NERSC) \cite{NERSC} of the Lawrence Berkeley National Laboratory.
In the latter case we used up to 1$\,$296 cores.

In the next section we report for both, PBC and DLT, the performance 
of our code as function of the number of CPU cores. This is followed 
by summary and conclusions. The appendix discusses annoying subtleties, 
which we encountered with MPI send and receive instructions, making 
slightly modified versions {\tt b} and {\tt c} of the code necessary 
for the test runs of this paper.

\section{Scaling with the number of processors} \label{Scaling}

At the FSU HEP group we connected two Intel E5405 2~GHz quad-core PCs 
by a crossover cable and installed Open MPI version 1.2.5-5. Up to 8 
MPI processes can be matched by the number of cores on one PC, up to 
16 on both PCs. The Fortran compiler used was g77 based on gcc version 
3.4.6 (Red Hat 3.4.6-4).

The Cray XT4 at NERSC features a configuration of 38$\,$640 AMD Opteron 
2.3~GHz quad cores with a SeaStar2 switch interconnect and MPICH2 
version 1.0.6p1 is installed using the Portland group compiler
version 8.0.1. On the Cray we have timed our code on up to 1$\,$296 
cores.

All CPU time measurements were done with the programs
\begin{equation} \label{su3time}  
  \tt cbsu3time2mpi\{a,b,c\}.f
\end{equation}
located in the folder {\tt ForProg} of the program package {\tt 
STMC2LSU3MPI} of Ref.~\cite{Be09}. These programs perform {\tt nequi} 
updating sweeps without measurements or read or write instructions. 
The value of {\tt nequi} is set in the parameter file {\tt mc.par}. 
Different versions {\tt a,b,c} are necessary to get MPI send
and receive instructions for all sublattice choices working,
see our appendix for details.

As there is no standardized Fortran time function, we rely on the Unix 
{\tt time} command to measure the execution time. CPU time needed
for initialization and creation of the start configuration was 
separately measured ({\tt lswp} false in {\tt mc.par}) and subtracted
when relevant.

\subsection{PC cluster}

The runs for this section are setup in the 
$$\tt 1TimeOpenMPI~~{\rm and}~~2TimeOpenMPI $$
folders of our {\tt STMC2LSU3MPI} tree.

\begin{table}[th] % BB Feb 10 2009. Collected in 1OPenMPI.
\centering
\caption{\label{tabPC}{Execution times on our cluster of two PCs for 
symmetric lattices at $\beta=5.7$ with {\tt mpifactor=1} and~2. The 
dimension of the MPI lattice is {\tt ndmpi}=n and the number of MPI 
processes np. The left part of the table is for PBC ({\tt nlat=1}) and 
the right part for DLT ({\tt nlat=2}). For PBC CPU times from a non-MPI 
run are listed in the third row.}}
\medskip
\begin{tabular}{|c|c|c|c|c|c|c|c|c|c|c|}
\hline
\multicolumn{2}{|c|}{Sweeps:}&512&101&32&2& &512&101&32&2  \\ \hline
 n &np&$8^4$&$12^4$&$16^4$& $32^4$&np& $8^4$&$12^4$&$16^4$&$32^4$\\ \hline
$-$&$-$ &48.2s& 48.3s& 48.7s& 48.9s&$-$&$-$ &$-$   &$-$    &$-$   \\ \hline
 1F&  1 &46.6s& 47.3s& 48.1s& 54.4s&$-$&$-$ &$-$   &$-$    &$-$   \\ \hline
 1 &  1 &46.9s& 47.7s& 48.5s& 54.5s& 2&48.4s& 48.2s&  48.9s& 54.5s\\ \hline
 1 &  2 &23.7s& 25.3s& 25.6s& 27.5s& 4&24.6s& 24.6s&  25.9s& 27.8s\\ \hline
 2 &  4 &12.5s& 12.3s& 12.7s& 14.0s& 8&13.0s& 13.3s&  13.2s& 13.5s\\ \hline
 3 &  8 & 7.1s&  6.7s&  6.8s&  6.9s&16&14.2s& 10.8s&   9.7s&  8.2s\\ \hline
 4 & 16 & 8.8s&  5.9s&  5.0s&  4.0s&$-$&$-$&$-$    &$-$    &$-$   \\ \hline
\end{tabular}
\end{table}

Table~\ref{tabPC} compiles CPU time measurements from runs of our program 
(\ref{su3time}) on our two PCs with 16 cores. As listed, lattice sizes 
are varied from $8^4$ to $32^4$. The left part of the table is for PBC 
({\tt nlat=1}), the right part for DLT ({\tt nlat=2}). Times for each 
lattice are normalized to the number of sweeps given in the first row 
of the table and taken inversely proportional to the lattice size. To 
get sufficiently accurate results, the actual numbers of sweeps were 
occasionally multiples of those given in the first row. Final CPU time 
uncertainties are about a few percent.

The first {\tt ndmpi} directions of the lattice are partitioned into 
sublattices. For {\tt mpifactor=2} the sublattice extensions in these 
directions are half of those of the entire lattice. As explained in 
\cite{Be09} these sublattices form a MPI lattice of size 
\begin{equation} \label{MPInp}
 \rm np = \tt msmpi = nlat * ndmpi**mpifactor\ .
\end{equation}
The number of MPI processes agrees with the number of points on the MPI 
lattice and each process is picked up by one core unless the number of 
processes exceeds the number of available cores.

For comparison we performed also 1-process CPU time measurements using
(a)~a non-MPI Fortran program indicated by $-$ in the first two columns
and (b)~our MPI code with {\tt mpifactor=1}, implying np=1. For the
1-process runs of our MPI program the parameter {\tt lbcex} of {\tt 
latmpi.par} allows one to turn the boundary exchange off or on, where 
off is indicated by a F in the nd column (for {\tt np}$\ge$2 boundary 
exchange has always be turned on). CPU times show that slowdown due 
to MPI send and receive instruction for boundary exchange is less than 
1\% when compared to the usual implementation \cite{Be04} of PBC by 
pointer arrays.

The other way round than on the PC used in \cite{Be09}, the non-MPI 
program is slightly slower than the MPI program.
%, can be attributed to its use of repeat until accepted heatbath updating, 
% while the MPI code uses the 1-hit version \cite{Be09}, which has 97.2\% 
% acceptance rate at $\beta=5.7$. 
For practical purposes the difference 
in CPU time consumption as well as in performance is negligible. For 
np=1 to~4 we notice some loss of performance on the $32^4$ system, 
which is likely due to inefficiencies of the used Fortran compiler. 
This disappears when the lattice is partitioned into small enough 
sublattices.

\begin{figure}[t]
 \begin{picture}(150,155)
    \put(0, 0){\includegraphics{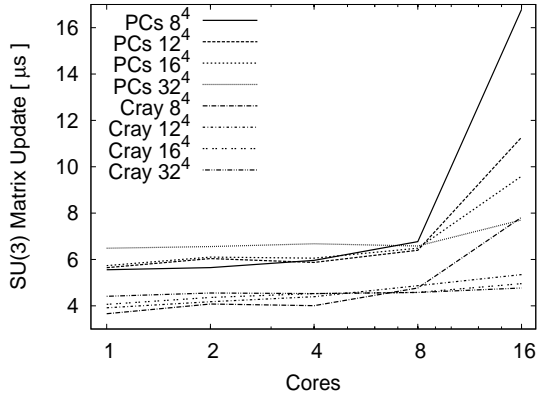}}
  \end{picture}
\caption{CPU time per SU(3) matrix update. \label{fig_CPU}}
\end{figure} % Generated in FiguresSTMC_SU3.

For PBC we use the CPU times in table~\ref{tabPC} and those from the 
Cray in table~\ref{tab1Cray}, to calculate update times per SU(3) 
matrix in units of single-core processor time. That is the execution
time multiplied with the number of cores and divided by the number
of SU(3) matrix updates performed. In Fig.~\ref{fig_CPU} we plot the
update times in microseconds $[\mu s]$ versus the numbers of cores 
used. Efficient performance is found in a range from 4 to 7 microseconds,
which is also typical for other SU(3) lattice gauge theory code like 
that of the MILC collaboration~\cite{MILC}, which is written in~C.
The reason for our better 1-processor performance on the Cray than
on the PCs appears to be that the Portland goup compiler is installed 
on the Cray, whereas on the PCs we had only the gnu compiler available.

On both, the PCs and the Cray, the CPU time per SU(3) update stays up 
to 8 cores almost constant, followed by a loss in efficiency when 16 
cores are employed. This loss is dramatic for small lattices on our 
PCs. Up to 8 cores we stay on one PC, while for 16 cores communication 
between the two PCs through the crossover cable is relatively slow. 
The surface to volume ratio of the employed sublattice matters then. This 
ratio is best (smallest) for the largest lattice. Compare $8\times 16^3
/16^4 = 0.5$ for the $32^4$ lattice to $8\times 4^3/4^4 = 2$ for the 
$8^4$ lattice. Due to more efficient communication between nodes, the 
effect is far more moderate on the Cray, although still visible.

The results reported in the right part of table~\ref{tabPC} are analogue 
to those of the left part. The main difference is that we now run on a 
DLT ({\tt nlat=2}). Scaling of CPU time with the number of cores is 
similar as before. The limit of 16 cores is already reached for {\tt 
ndmpi=3}. 

Using a number of MPI processes, which exceed the number of cores
is possible but inefficient. Running an $8^4$ lattice with 16 MPI 
processes on one PC with 8 cores needs 30\% more CPU time than running 
the same job with 8 MPI processes. Partitioning a $24^3$ lattice with 
{\tt mpifactor=3} we found for {\tt ndmpi=1}, i.e. 3 MPI processes, an 
improvement factor of 2.4 in real time compared to the 1-process run. 
For a run with 9 MPI processes the further improvement factor was only 
1.4 compared to the 3-processes run. Note also that one should not 
execute other jobs in the background even with {\tt nice~19}. Running 
on one PC 8 additional jobs with {\tt nice~19} took only 5\% of the CPU 
time according to the information provided by the {\tt top} command.
But due to the resulting uneven balancing the execution time of a MPI 
with 8 processes went actually up by 35\% in real time (while getting
95\% of the CPU time).

\subsection{Cray}

\begin{table}[th] % BB Feb 9 2009. Collected in 1OPenMPI.
\centering
\caption{\label{tab1Cray}{Runs on the Cray analogue to those 
of table~\ref{tabPC}.}} \medskip
\begin{tabular}{|c|c|c|c|c|c|c|c|c|c|c|c|c|} \hline
\multicolumn{2}{|c|}{Sweeps:}&512&101&32&2& &512&101&32&2\\ \hline
 n &np &$8^4$&$12^4$&$16^4$&$32^4$&np &$8^4$&$12^4$&$16^4$&$32^4$\\ \hline
 1 & 1F&30.7s&32.8s&34.1s& 37.0s&$-$& $-$& $-$  & $-$&$-$\\ \hline
 1 & 1 &32.7s&33.7s&34.7s& 37.4s&  2&34.4s&34.3s& 35.4s&41.2s\\ \hline
 1 & 2 &17.1s&17.5s&18.3s& 19.1s&  4&16.9s&17.6s& 18.3s&20.5s\\ \hline
 2 & 4 & 8.4s& 9.2s& 9.5s&  9.5s&  8& 8.5s& 9.2s&  9.6s& 9.6s\\ \hline
 3 & 8 & 5.0s& 5.1s& 4.8s&  4.8s& 16& 5.3s& 5.2s&  4.8s& 4.9s\\ \hline
 4 &16 & 4.1s& 2.8s& 2.6s&  2.5s& 32& 4.5s& 3.2s&  2.7s& 2.5s\\ \hline
\end{tabular} \end{table}

Examples of jobs for this section are setup in 
$$\tt 1CrayTime~~{\rm and}~~2CrayTime $$
folders of our {\tt STMC2LSU3MPI} tree. They are not as easily
reproducable as our previously discussed runs, because a 
supercomputer is needed, which will rely on its particular 
job control commands (those for the NERSC Cray are in the files
{\tt q.run*}, the ouput in {\tt *.out}).

\begin{figure}[t] \begin{picture}(150,155)
    \put(0, 0){\includegraphics{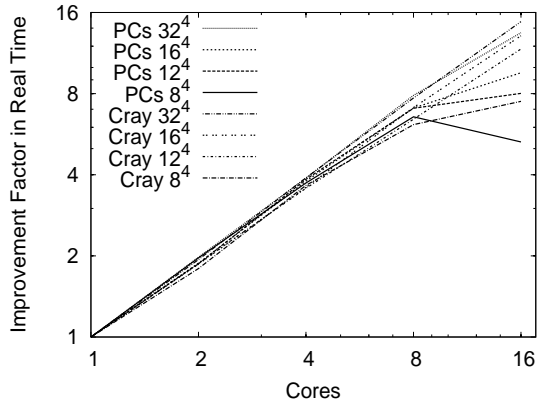}}
  \end{picture}
\caption{Improvement factors in real time. \label{fig_FCT}}
\end{figure} % Generated in FiguresSTMC_SU3.

Table~\ref{tab1Cray} compiles CPU time measurements on the Cray,
which are analogue to those of table~\ref{tabPC}. 
In Fig.~\ref{fig_FCT} 
we use the results of tables~\ref{tabPC} and~\ref{tab1Cray} to plot the 
improvement factor in real time, defined as
\begin{equation} \label{FCT}
  \rm FCT = time(1)/time(np)\,, 
\end{equation}
versus the number of cores used. Here time(1) is the CPU needed without 
parallelization and time(np) the CPU time per core for running with np 
processes on np cores. The real time one has to wait for completion of 
a job is inversely proportional to~FCT.

Up to 8 cores the relationship (\ref{FCT}) is practically linear in the
number of cores. For the $16^4$ lattice the slope is 0.88 on the PCs as 
well as on the Cray, but in the range from 8 to 16 cores it is 0.75 on 
the Cray and down to 0.3 on the PCs. For the $32^4$ lattice the slope 
is above 0.95 for up to 8 cores on the PCs as well as on the Cray. 
Then, in the range from 8 to 16 cores it drops to 0.89 on the Cray 
and to 0.70 on the PCs.  Parallelization beyond 8 cores on the PCs 
makes no sense on the $8^4$ lattice, for which the slope between 8 
and 16 cores is negative.  

\begin{table}[tb] % BB Feb 9 2009. Collected in 1OPenMPI.
\centering
\caption{\label{tab2Cray}{Execution times for 256 sweeps at 
$\beta=5.7$ with PBC ({\tt ndmpi=4}, left) and DLT ({\tt ndmpi=3}, 
right), nf={\tt mpifactor} in both cases and np number of processes.}} 
\medskip
\begin{tabular}{|c|c|c|c|c|c|c|} \hline
nf & np &Lattice&Time  & np & Lattice  & Time  \\ \hline
 1 &   1&  $8^4$& 16.7s&   2&  $8^3\,8$& 17.7s \\ \hline
 2 &  16& $16^4$& 20.4s&  16& $16^3\,8$& 19.6s \\ \hline
 3 &  81& $24^4$& 22.3s&  54& $24^3\,8$& 20.4s \\ \hline
 4 & 256& $32^4$& 23.2s& 128& $32^3\,8$& 20.6s \\ \hline
 5 & 625& $40^4$& 23.9s& 250& $40^3\,8$& 21.1s \\ \hline
 6 &1296& $48^4$& 28.2s& 432& $48^3\,8$& 21.3s \\ \hline
 7 &$-$ & $$-$ $&  $-$ & 686& $56^3\,8$& 22.6s \\ \hline
 8 &$-$ & $$-$ $&  $-$ &1024& $64^3\,8$& 24.3s \\ \hline
\end{tabular} \end{table}

Ultimately, one wants to employ many cores for runs on large lattices.
In table~\ref{tab2Cray} we use $8^4$ sublattices for parallelization 
of lattices with PBC up to size $48^4$ ({\tt ndmpi=4} and {\tt 
mpifactor=6}) and for DLT lattices up to size $64^38$ ({\tt ndmpi=3}
and {\tt mpifactor=8}).

\begin{figure}[t] \begin{picture}(150,155)
    \put(0, 0){\includegraphics{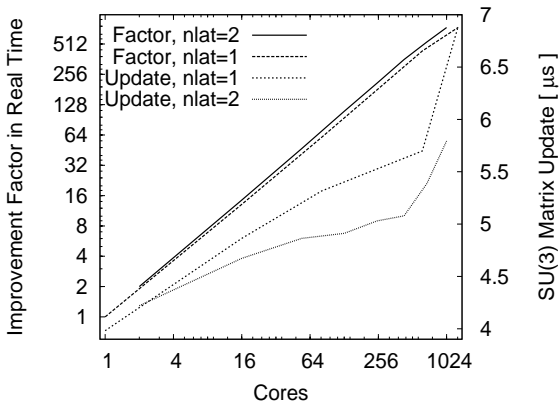}}
  \end{picture}
\caption{Cray runs with up to 1$\,$296 cores. \label{fig_Cray}}
\end{figure} % Generated in FiguresSTMC_SU3.

Figure~\ref{fig_Cray} plots as function of up to 1$\,$296 cores the 
update times of SU(3) matrices, given on the right ordinate, together 
with the improvement factors in real time, given on the left ordinate.
All update times stay below 7~$\!\mu$s.
The log scale on the left hides to some extent that the performance 
on the DLT is better than with PBC (note, however, that we used {\tt 
ndmpi=3} on the DLT and {\tt ndmpi=4} with PBC). Measured in 
percentages of the peak speed obtained in single processor runs one 
finds up to 432 cores an efficiency of about 83\% for the DLT.
For PBC in the range of 256 to 625 cores it is around 70\%.
With 1$\,$296 cores the performance with PBC drops to 58\%, while 
it is still 73\% when simulating the DLT with 1$\,$024 cores. In 
all cases there is a considerable gain in real time when increasing 
the number of participating cores. Users are advised to tune {\tt 
ndmpi} and {\tt mpifactor} for optimal performance on a particular 
supercomputer before running large scale production, evaluating then 
also the performance of measurement routines.

\section{Summary and Conclusions} \label{Conclusions}

As computer configurations with large numbers of processors drop in 
price, while the peak performance of single CPUs is almost stagnant, 
non-trivial parallel processing becomes more important than ever. When 
running parallel applications, the user is then first of all interested 
in his or her gain in real (wait) time. So, let us conclude with 
examples from runing our MPI code. For our $32^4$ lattice on a cluster 
of two PCs the reduction in real time is by a factor 1/7.88 when 
running on eight cores and by 1/13.5 when running on all 16 cores. The 
corresponding factors on the Cray are 1/7.7 (down due to better single 
core performance) and 1/14.8 (up due to better networking). Scaling 
$8^4$ sublattices on a DLT up to size $64^38$ and using 1$\,$024 cores 
on the Cray, the reduction in real time is by a factor 1/746. Scaling 
$8^4$ sublattices to a $48^4$ lattice with PBC and using 1$\,$296 cores 
it was 1/749. 

\bigskip
\noindent {\bf Acknowledgments:}
This work was in part supported by the U.S. Department of Energy under 
contract DE-FG02-97ER41022 and by the German Humboldt Foundation. 
BB thanks Wolfhard Janke, Elmar Bittner and other members of the 
Institute for Theoretical Physics of Leipzig University for their 
kind hospitality.

\appendix
\section{MPI Send and Receive Subtleties} 

The problems discussed in this appendix are related to allocating enough 
buffer space, so that MPI can send and receive arrays of a requested 
size. In our 4D runs the size of the sublattice boundary arrays, which 
MPI has to send and receive, is given by the size of the checkerboard 
SU(3) matrix storage perpendicular to the 1-direction:
\begin{equation} \label{mscb18}
   \tt 18 * mscb,\ mscb = nl1*nl3*nl4/2\ .
\end{equation}
For our Open MPI implementation at FSU the maximum size for a Real*8 
array, which could be transferred using basic MPI send and receive 
instructions turned out to be 503, even too small to run our SU(3) 
MPI program with $4^4$ sublattices. For the MPICH installation at the 
Institute for Theoretical Physics of Leipzig University the number 
turned out to be 15$\,$999 and on the NERSC Cray with MPICH2 it was 
16$\,$384. While the latter numbers are sufficiently large to allow 
for most applications, there are exceptions. For instance, only the 
last ({\tt ndmpi=4}) of the $32^4$ lattice runs of table~\ref{tab1Cray} 
is possible.

We were unable to find documentation of these array size limits. When 
the program tries to send an array larger than the allowed maximum 
size, one encounters a hangup without any error messages. Therefore, 
before submitting a MCMC job in its version {\tt a}, we recommend to 
check that the array size (\ref{mscb18}) can really be transmitted. For 
this purpose the program
\begin{equation} \label{send}
  {\tt dsenda.f}
\end{equation}
is kept in {\tt MPICHtest} and {\tt OpenMPItest} subfolders of the
{\tt 1MPICH} and {\tt 1OpenMPI} projects. In this program we have set 
the array size parameter {\tt NDAT} to the value for which the array 
transmission works on our platforms, while it fails when increasing 
the initial {\tt NDAT} value by $+1$. Also included is a corresponding 
program, {\tt isenda.f}, which tests on integer arrays.

Before submitting a SU(3) MPI job: Change the {\tt NDAT} parameter 
in the {\tt dsenda.f} program to the array size, which you need, and 
confirm that the array transfer works on your MPI platform. If yes, 
you can use the {\tt a}-version of our SU(3) programs. If the array 
transfer hangs up, you will need a solution similar, but not 
necessarily identical, to those given in our {\tt b} and {\tt c} 
versions.

For our Open MPI at FSU it was possible to overcome the buffer problem 
by modifications, which are given in the following. Instead of the main 
program listed in Ref.~\cite{Be09} the version
\begin{equation} \label{SU3mainb}
  {\tt cbsu3\_dltb.f} 
\end{equation}
({\tt b} for ``buffer'') has to be used. It replaces 
in the following four include statements \smallskip
\begin{tiny} \begin{verbatim}
      include '../../Libs/MPISU3/cbsu3_bnd1a_mpia.f' ! Gather boundary.
      include '../../Libs/MPISU3/cbsu3_bnd1b_mpia.f' ! Gather boundary.
      include '../../Libs/MPISU3/cbsu3_bnd2a_mpia.f' ! Gather boundary.
      include '../../Libs/MPISU3/cbsu3_bnd2b_mpia.f' ! Gather boundary.
\end{verbatim} \end{tiny} \smallskip
{\tt ...mpia.f} by {\tt ...mpib.f}. This exchanges the routines, which
perform the crucial MPI send and receive instructions. The plain version 
is replaced by a routine using buffered send and receive instructions. 
The parts of the two subroutines, where the relevant differences lie, 
are listed in the following. In {\tt cbsu3\_bnd1a\_mpia.f} it is \smallskip

\input CODE/cbsu3_bnd1a_mpia.f \smallskip

\noindent compared with {\tt cbsu3\_bnd1a\_mpib.f}: 
\smallskip

\input CODE/cbsu3_bnd1a_mpib.f \smallskip

\noindent In the {\tt b} version 
$$ {\tt mpi\_buffer\_attach}\,,~~ {\tt mpi\_buffer\_detach} $$ 
statements have been added, and {\tt mpi\_send} has been replaced by 
{\tt mpi\_bsend}. Note that the Fortran names of the {\tt cbsu3\_bnd*.f}
routines have been kept (in contrast to their filenames), so that one 
has only to exchange the include statements in the main program. 

Unfortunately, the buffered send and receive instructions of our {\tt 
b} version are not universal MPI code. They neither work with the 
MPICH installation at Leipzig University nor with MPICH2 on the NERSC
Cray. Extended buffer sizes in our version {\tt c} program,
\begin{equation} \label{SU3mainc}
  {\tt cbsu3time2mpic.f}\ ,
\end{equation}
which is included in the {\tt Cray} project folders of our package,
performed well with MPICH2 on the Cray, but failed with Open MPI at
FSU and MPICH at Leipzig University. A unified MPI standard appears
to be missing.

\end{document}